\documentclass[final]{svjour3}
\usepackage{graphicx}
\usepackage{rotating}
\usepackage{amssymb}
\usepackage{mathptmx}
\makeatletter
\journalname{Journal of Low Temperature Physics}

\begin{document}

\authorrunning{Austermann et al.}
\titlerunning{Millimeter-Wave Polarimeters Using Kinetic Inductance Detectors for TolTEC and Beyond}

\newcommand{\hdblarrow}{H\makebox[0.9ex][l]{$\downdownarrows$}-}
\title{Millimeter-Wave Polarimeters Using Kinetic Inductance Detectors for TolTEC and Beyond}

\author{J.E.~Austermann$^1$ \and J.A.~Beall$^1$ \and S.A.~Bryan$^2$ \and B.~Dober$^1$ \and J.~Gao$^1$ \and G.~Hilton$^1$ \and J.~Hubmayr$^1$ \and P.~Mauskopf$^2$ \and
C.M.~McKenney$^1$ \and S.M.~Simon$^3$ \and J.N.~Ullom$^1$ \and M.R.~Vissers$^1$ \and G.W.~Wilson$^4$}

\institute{
1: Quantum Sensors Group, National Institute of Standards and Technology; Boulder, CO 80305, USA\\ 
2: Arizona State University, Tempe, AZ\\
3: University of Michigan, Ann Arbor, MI\\
4: University of Massachusetts, Amherst, MA\\
\email{jea@nist.gov}}

\maketitle

\begin{abstract}

Microwave Kinetic Inductance Detectors (MKIDs) provide a compelling path forward to the large-format polarimeter, imaging, and spectrometer arrays needed for next-generation experiments in millimeter-wave cosmology and astronomy.   We describe the development of feedhorn-coupled MKID detectors for the TolTEC millimeter-wave imaging polarimeter being constructed for the 50-meter Large Millimeter Telescope (LMT).  Observations with TolTEC are planned to begin in early 2019.  TolTEC will comprise $\sim$7,000 polarization sensitive MKIDs and will represent the first MKID arrays fabricated and deployed on monolithic 150~mm diameter silicon wafers -- a critical step towards future large-scale experiments with over $10^5$ detectors.  TolTEC will operate in observational bands at 1.1, 1.4, and 2.0 mm and will use dichroic filters to define a physically independent focal plane for each passband, thus allowing the polarimeters to use simple, direct-absorption inductive structures that are impedance matched to incident radiation.   This work is part of a larger program at NIST-Boulder to develop MKID-based detector technologies for use over a wide range of photon energies spanning millimeter-waves to X-rays.  
We present the detailed pixel layout and describe the methods, tools, 
and flexible design parameters that allow this solution to be optimized for use anywhere in the millimeter and sub-millimeter bands.
We also present measurements of prototype devices operating in the 1.1~mm band and compare the observed optical performance to that predicted from models and simulations.


\end{abstract}

\section{Introduction}

\begin{figure}[t]
\begin{center}
\includegraphics[width=1.0\linewidth,keepaspectratio]{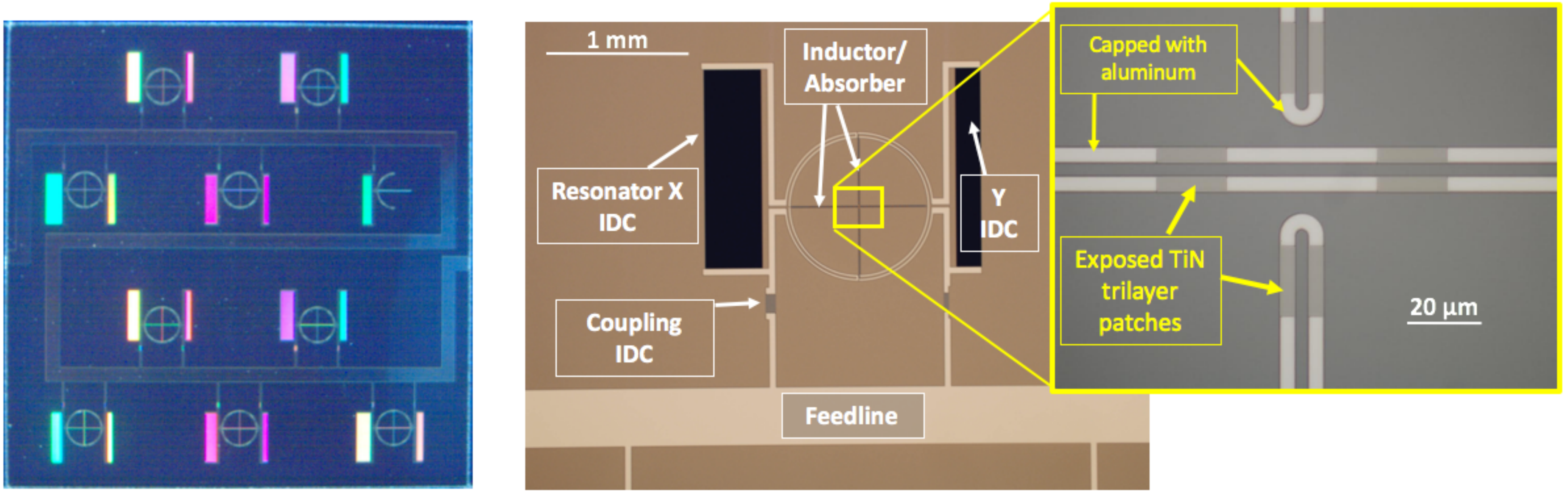}
\caption{{\it Left:} Photograph of a 15~mm $\times$ 15~mm prototype array of 10 pixels for the 1.1~mm band.
{\it Center:} Photograph depicting the layout of a single TolTEC dual-polarization pixel.
{\it Right:} Expanded view of the central region of the inductors/absorbers highlighting the use of aluminum shorting patches.
}
\end{center}
\label{fig1}
\end{figure}

Since being described by Day et al. in 2003~\cite{day03}, Microwave Kinetic Inductance Detectors (MKIDs) have rapidly advanced so that they are now
a viable and often compelling technological choice for the principal sensor in broadband imagers~\cite{golwala12a,Calvo16}, polarimeters (e.g. BLAST-TNG~\cite{hubmayr15a,dober15}), and spectrometers (e.g. \cite{shirokoff14,mazin13}) over a wide range of the electromagnetic spectrum.
The chief attraction of KIDs is the ease with which large sensor arrays can be assembled and read out using GHz frequency division multiplexing with minimal interconnects. This allows for many hundreds to several thousands of channels to be measured on a single pair of coaxial cables (e.g. \cite{bourrion16,galitzki14}) with simple assembly and integration.  
For some KID architectures, such as that outlined here, fabrication is also simpler and faster than competing approaches.  

In this report, we describe the design and performance of millimeter-wave, feedhorn-coupled, direct-absorption, polarization-sensitive pixels using microwave kinetic 
inductance detectors (MKIDs).
This research is a direct extension of the NIST-Boulder sub-millimeter MKID polarimeters developed for BLAST-TNG~\cite{hubmayr15a,dober15}.
These millimeter-wave versions are being developed for TolTEC 
-- a new imaging polarimeter for the 50-meter 
Large Millimeter Telescope (LMT) -- 
and other future experiments in astronomy and cosmology. 
TolTEC is a three color millimeter-wave polarimeter designed to perform a series of large legacy surveys that will address many of the fundamental questions related to the formation and evolution of structure on scales from stars to clusters of galaxies.
In total, TolTEC will consist of approximately 7,000 polarization sensitive MKIDs across three focal planes. 
In this work, we describe a scalable polarimeter design for operation at millimeter and sub-mm wavelengths.
In particular, we present results for a dual-polarization pixel operating in a band centered at $\sim$1.1~mm and compare the measured optical performance to that expected from detailed simulation and models.  

\section{Polarimeter Design}

TolTEC will use dichroic filters~\cite{ade06} to define a physically independent focal plane for each of three passbands centered at 1.1~mm, 1.4~mm, and 2.0~mm.  
The single-band nature of each focal plane allows the polarimeters to use simple, direct-absorption resistive structures that are impedance matched to incident radiation through a feedhorn coupled waveguide.
The detector's bandpass is primarily defined through a combination of metal-mesh, free-space, low-pass filters~\cite{ade06} and high-pass dichroics, as well as the cutoff frequency of a section of circular waveguide at the detector end of each feedhorn.
Similar to the BLAST detectors, the detector arrays are front-side illuminated and 
reflective quarter-wave backshorts are created by depositing the detectors on
a silicon-on-insulator (SOI) substrate with a quarter wave device layer thickness and depositing an aluminum ground plane on the wafer backside (see \cite{hubmayr15a}).
This approach allows for a simple photolithographic fabrication process with 
a single wiring layer and no electrical crossovers.

\begin{figure}[t]
\begin{center}
\includegraphics[width=0.95\linewidth, keepaspectratio]{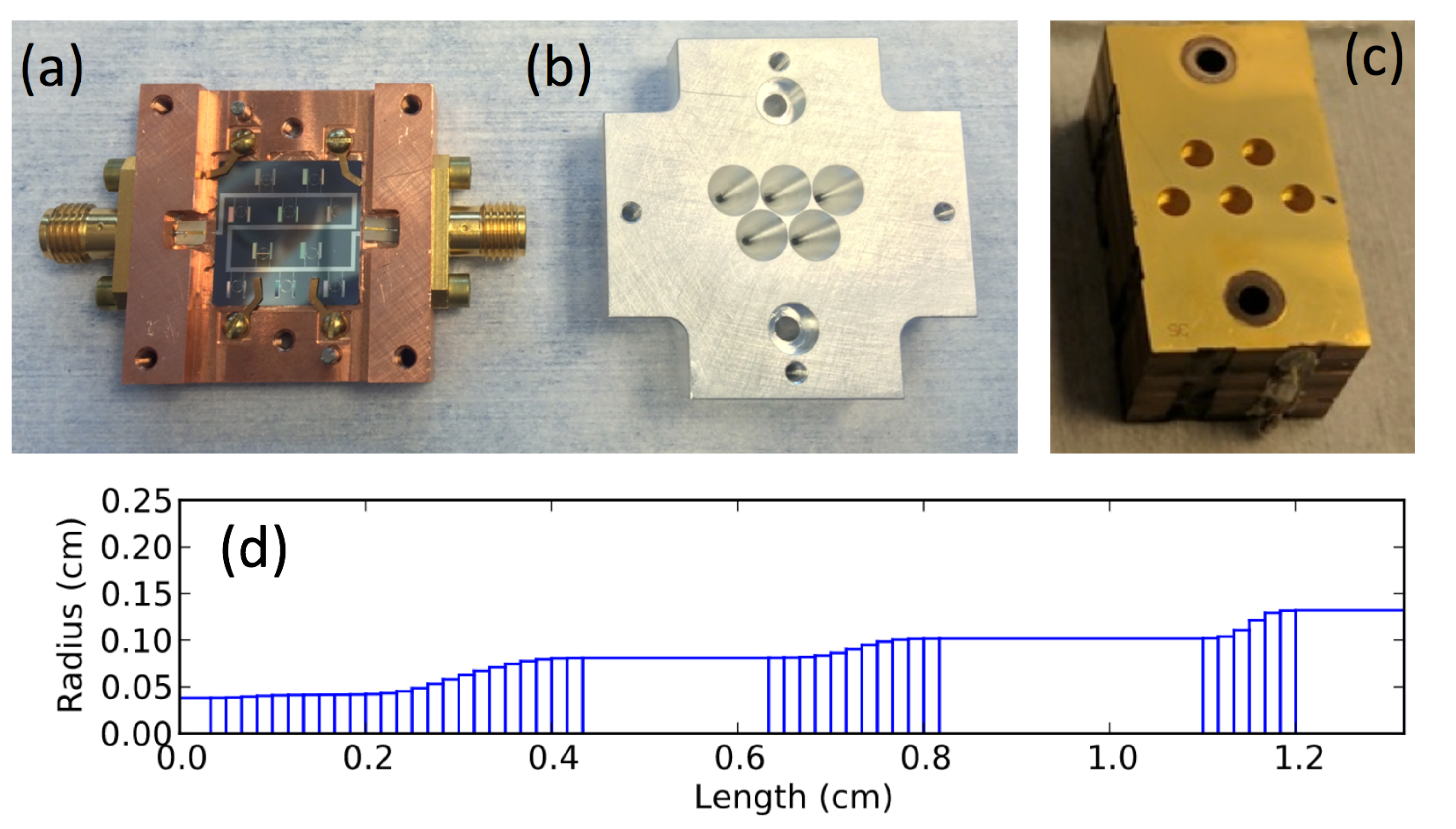}
\caption{{\it (a):} Experimental mount with prototype sub-array of 1.1~mm polarimeters. {\it (b):} Matching metal feedhorn array for lab measurement of central 5 pixels.
{\it (c):} Silicon platelet feedhorn sub-array prototype {\it (d):} Silicon feedhorn radial profile.  
Vertical lines represent a change in hole radius at that position.  Individual wafers of thickness 500~$\rm{\mu}$m or 333~$\rm{\mu}$m
(optionally double-etched for an effective 167~$\rm{\mu}$m step size) are used to create the profile.
}
\end{center}
\label{fig2}
\end{figure}

TolTEC pixels comprise two lumped-element MKIDs that are sensitive to orthogonal linear polarization states of incident radiation.
Each detector forms a resonant microwave circuit consisting of a 4~$\mu$m wide inductive strip and an interdigitated capacitor (IDC) designed to resonate at
a unique frequency (see Fig~\ref{fig1}).
When producing large networks of resonators, we typically design the inductor to be identical in every pixel while each IDC is trimmed to a unique capacitance using
the stepper lithography techniques described in~\cite{liu17b,mckenney18submit}.
The inductive strips are made of proximitized TiN/Ti multilayers, which have a tunable transition temperature, $T_{\rm{c}}$, that is highly uniform across 
an array when compared to sub-stoichiometric TiN~\cite{vissers13b}.   
This material also has a tunable 
sheet resistance, $R_{\rm{s}}$, through the choice of TiN and Ti layer thicknesses and the number of layers.  
For the 1.1~mm prototype devices discussed here, we have used a TiN/Ti/TiN trilayer with thicknesses of 4/10/4~nm, respectively,
resulting in $T_{\rm{c}} = $1.4~K, $R_{\rm{s}} \sim 80$~Ohm/$\square$, and inductance of $L_{\rm{s}} \sim 90$~pH/$\square$.
We have found that a TiN-silicon substrate interface exhibits low two-level system (TLS) noise 
\cite{vissers13b,vissers10a}, allowing TolTEC to utilize the compact IDCs
necessary to achieve the instrument design of $1f\lambda$ pixel spacing ($\sim$2.75~mm pixel-to-pixel spacing for 1.1~mm band) with resonance frequencies in the readout band of 0.5--1.0~GHz. 

The TiN/Ti/TiN trilayer serves both as the inductor of the resonant circuit and the absorber that couples to radiation from the waveguide.  
To accomplish both these tasks with optimal performance, the multilayer geometric design must balance the following factors:  (a) an appropriate transition temperature for observed photon energies (determined by TiN thickness); 
(b) polarization efficiency (a function of absorber trace width); 
(c) detector responsivity and saturation power (a function of multilayer volume); 
and (d) be optimally impedance matched to the incident radiation (accomplished through tuning of R$_{\rm{s}}$ and inductor geometry).
An optimal solution to these 4 factors requires an additional design parameter beyond the basic 3 geometric dimensions of the absorber. 
We accomplish this by separating the duties of impedance matching and TiN geometry by depositing a 100~nm thick layer of aluminum over portions of the inductor/absorber strip (Fig~\ref{fig1}). 
With significantly lower R$_{\rm{s}}$ and L$_{\rm{s}}$ than the TiN multilayer, the Al acts as a short that adds minimal impedance and inductance to the overall circuit.  
This allows the absorbing strip to be designed with almost any volume of inductor while maintaining optimal coupling to the waveguide with high polarization efficiency.
In practice, the inductor and aluminum patch geometries are optimized over the observation band through simulations of a parameterized model using finite element simulations while holding the desired multilayer volume and inductance constant.

Simulations suggest $\sim5\%$ of incident radiation is directly absorbed by the low-inductance aluminum shorts in these prototypes. 
This represents a small reduction in optical efficiency that can be improved in future designs through the use of shorting material and/or geometries with lower impedance.
The shorts can also affect detector responsivity through the diffusion of quasiparticles from the higher gap energy trilayer ($T_{\rm{c}}\sim1.4$~K) to the lower gap aluminum ($T_{\rm{c}}\sim1.2$~K).  
We have measured devices with varying lengths of exposed absorber which show that trilayer patch lengths of $\gtrsim 15$~$\mu$m will retain $\gtrsim$~50\% of the responsivity expected in a continuous absorber of equivalent volume and kinetic inductance fraction. 
For TolTEC, the nominal trilayer patch is 25~$\mu$m long and each detector has a total active trilayer volume of 26~$\mu$m$^3$, resulting in a responsivity (Sec.~\ref{sec3}) that allows the detector to operate well within the photon-limited regime under the expected photon load, and thus quasiparticle diffusion will have a negligible effect on detector sensitivity. 
For future applications that may require significantly shorter trilayer patch lengths, we are exploring tuning $T_{\rm{c}}$ values of the absorber and/or shorting materials such that the the shorts have a higher gap energy, i.e. ${\Delta_{\rm{short}}} > {\Delta_{\rm{abs}}}$, which would lead to significantly reduced quasiparticle diffusion rates.

 
\begin{figure}[t]
\begin{center}
\includegraphics[width=1.0\linewidth, keepaspectratio]{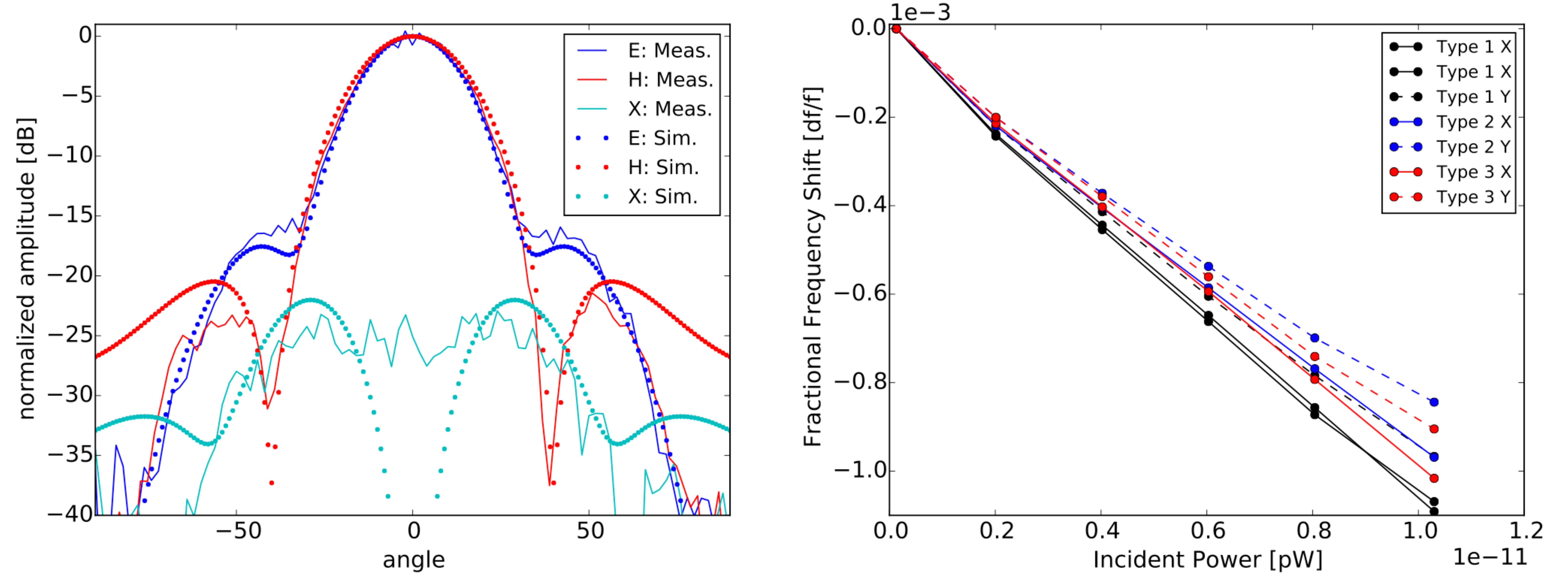}
\caption{{\it (Left)}:  Feedhorn-only measurements of angular response (beam) of the 1.1~mm band silicon platelet feedhorn stack at 255~GHz. E-plane, H-plane, and X-pol (cross-polarization) measurements are compared to simulation.
Differences between measurement and simulation at high angles in the H-plane are due to an occulting mounting structure inherent in the measurement setup.
The predicted X-pol dip at the center of the beam is typically difficult to measure within the alignment tolerance of the experimental setup.
{\it (Right):} Responsivity of the X and Y oriented detectors of three pixel design variations when coupled to a temperature controlled black body varied between 3~K and 20~K.  The small systematic difference between the X and Y channels of each pixel could be due to several factors
including inductor and capacitor geometric differences in design,
differences in optical efficiency, and waveguide misalignment and/or ellipticity.  
}
\label{fig3}
\end{center}
\end{figure}

\section{Optical Performance}
\label{sec3}

Prototype devices for the 1.1~mm band have been fabricated as 10-pixel arrays 
(see Fig.~\ref{fig1}).
Several pixel geometries are included on each array in order to measure the performance of varying designs.
The resonators have been designed to operate within the nominal TolTEC readout band (0.5--1.0~GHz) and 
are measured to have an optically dark internal quality factor, $Q_{\rm{i}}$, of $2-3\times10^{5}$ at bath temperatures $T_{\rm{bath}} \lesssim 200$~mK
(TolTEC operates at $\sim$~100~mK).

Five of the pixels on the array are optically coupled through aluminum conical feedhorns (Fig~\ref{fig2}) that are oversized in order to match the laboratory optics used in the measurements described in this section.  
The deployed version of TolTEC pixels will be coupled using spline-profiled silicon platelet feedhorns (Fig.~\ref{fig2}) that have been numerically optimized
to maximize aperture efficiency while maintaining beam symmetry and polarization performance \cite{simon16}.  
We have fabricated a prototype 5-pixel array of these silicon horns and 
measured the beam response using a room temperature vector network analyzer.
The beam measurements of the prototypes are well matched to the simulations (e.g. Fig.~\ref{fig3}), 
validating our model and allowing us to proceed with the full production of the final feedhorn array consisting of $\sim$40 etched 150~mm diameter wafers.   
The detector beams will be truncated with a -3~dB edge taper in the final TolTEC instrument using a 4~K Lyot-stop.

\begin{figure}[t]
\begin{center}
\includegraphics[width=1.0\linewidth, keepaspectratio]{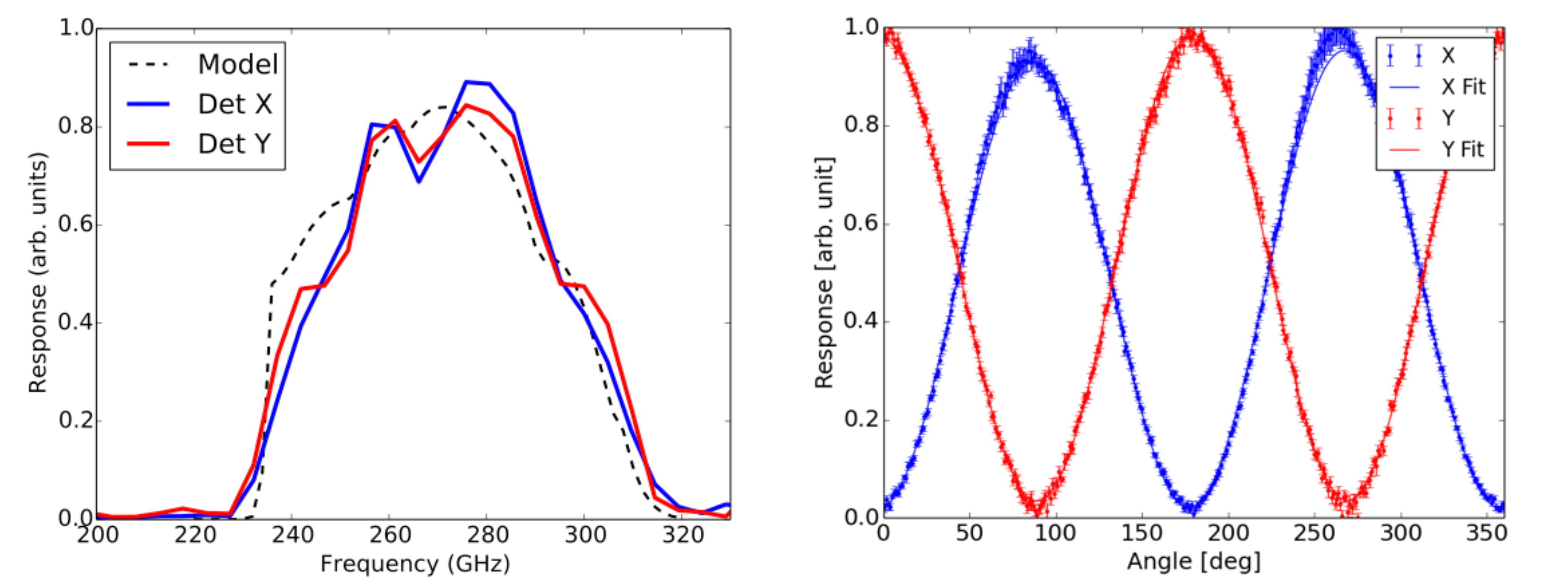}
\caption{{\it (Left)}: Measured passband of the X and Y polarization detectors of a single pixel compared to a transmission model constructed from the simulated waveguide-coupled detector efficiency and the expected transmission of low-pass filters used for this measurement.  
Measurements are normalized to have the same band integrated power as the model.
{\it (Right):} Normalized response as a function of position angle of a linearly polarized source.  The data has been fit to a model (solid lines) consisting of detector co-polar angle and amplitude, cross-pol amplitude, and a 180~degree source asymmetry (e.g. misalignment of rotating source).  
}
\label{fig4}
\end{center}
\end{figure}

We measure the detector response to incident radiation (Fig.~\ref{fig3}, {\it{right}}) by coupling the radiation from a temperature controlled black body through feedhorn-coupled waveguide (Fig.~\ref{fig2}).
Measurements are made at various black-body temperatures in the range of 3--20~K. 
Black-body temperature is converted to incident power on the detector using a passband model outlined in Fig.~\ref{fig4}.
Measurements of detector noise are used to estimate the optical efficiency as outlined in \cite{yates11,hubmayr15a}.
All three pixel design variations display photon-noise limited performance at
the expected loading ($\sim$10~pW) with measured optical efficiency in the range 71\%--80\%, which is consistent with both the simulation prediction of 80\% and the efficiency of BLAST-TNG devices of similar design \cite{dober15}.

Additional detector characterization is conducted by optically coupling the detector to room temperature experiments.  
This is accomplished through a series of optical elements within the cryostat, 
including a Zotefoam PPA30\footnote{http://zotefoams.com} vacuum window, a series of 
metal-mesh low-pass filters \cite{ade06}, and a free-space optical attenuator made of
Eccosorb MF-110\footnote{http://www.eccosorb.com/} with a thickness
that results in an average of $\sim5\%$ transmission in the 1.1~mm band.

Passbands of X and Y detectors were measured using a Fourier-transform spectrometer (FTS) and 
are shown in Fig.~\ref{fig4}.  
The X/Y channels appear to be well matched to each other and 
are consistent with the designed band edges.  
The passband shape is primarily dictated by the high-pass cutoff frequency of the
waveguide and the free-space low-pass filters used along the line of sight.
We note that the expected passband differs slightly from what is designed for the final TolTEC experiment (roughly 245--310~GHz) in a few ways.  
First, these laboratory measurements used low-pass filters of a different cut-off frequency due to filter availability at time of measurement.
Second, the low-frequency edge of the TolTEC band will be primarily defined by a dichroic filter rather than the waveguide. 
The waveguide cutoff frequency is designed to be below the nominal band edge in order to be sensitive to the full range of the dichroic's relatively shallow transition from reflective to transmissive as a function of frequency. 

We also perform polarization calibration by measuring the detector response to 
a polarized source -- a chopped black body behind a rotating wire grid --
as a function of grid polarization angle (Fig~\ref{fig4}). 
Model fits to these data show the X and Y detectors to be sensitive to orthogonal linear
polarization states with total (detector + optics) 
cross-polar leakage of $2.3\pm0.1\%$ and $3.6\pm 0.1\%$ (stat.-only uncertainties) 
for the X and Y channels, respectively.  
These measurements are roughly consistent with the simulated prediction of $\sim2\%$ 
cross-polar leakage from the detectors.

\section{Ongoing Development}

TolTEC is currently scheduled to achieve first light at the 50-meter LMT in early 2019 with the operating parameters listed in Table~\ref{tab:basics}.
To this end, we continue to characterize and optimize the 1.1~mm pixels and are concurrently developing and optimizing designs for the 1.4~mm and 2.0~mm bands.  
We are also moving towards the production of large-format detector arrays on 150~mm diameter substrates that are required for the final instrument.  We have now produced an array of over 2000 resonators on a 150 mm substrate in order to characterize device uniformity over these large scales.  For the deployed arrays, we aim to implement a recently developed LED mapper and 
post-measurement lithographic capacitor correction
technique in order to achieve highly uniform frequency spacing between resonators and detector+readout yield near $100\%$ (see \cite{liu17b}).

The measurements presented here, together with results from similar
BLAST-TNG devices optimized for sub-mm wavelengths \cite{hubmayr15a,dober15}, show that materials and pixel designs are now available that meet the needs of various experiments in the sub-mm and mm-wave regime.  The successfully predictive power of our models and simulations allow for rapid design optimization. Together, these capabilities help to make large KID arrays an attractive and realistic sensor choice over a wide range of observation wavelengths.  

\begin{table}[t]
\small
\begin{center}
\begin{tabular}[c]{ccccc}
\hline
\bf{Notional Band Center} & \bf{1.1~mm} & \bf{1.4~mm} & \bf{2.0~mm} \\
\hline
Target Frequency Band   & 245--310~GHz     &  195--245~GHz   &  128--170~GHz  \\
Approx. Number of Detectors  & 3800     &          1900         & 950   \\
Expected Loading (median)   &  10.7~pW      &  7.2~pW           & 4.8~pW   \\
\hline
\end{tabular}
\caption{Targeted operational parameters for the final TolTEC detector arrays, with two detectors per pixel.  Per detector loading is calculated using an atmospheric model for the LMT site at median opacity during the observing season, as well as estimates of emitted and scattered light from all optical elements. 
}
\label{tab:basics}
\end{center}\vskip -0.2 in
\end{table}

\begin{acknowledgements}
This work is supported by the NSF of the United States through grant MSIP-1636621 and NASA through grant APRA13-0083.
\end{acknowledgements}


\bibliographystyle{aipprocl}
\bibliography{Bibliography}

%

\end{document}